# New Approaches to Final Cooling

David Neuffer

*Fermilab*\*, *PO Box 500 Batavia IL 60510*

**Abstract.** A high-energy muon collider scenario require a "final cooling" system that reduces transverse emittance by a factor of ~10 while allowing longitudinal emittance increase. The baseline approach has low-energy transverse cooling within high-field solenoids, with strong longitudinal heating. This approach and its recent simulation are discussed. Alternative approaches which more explicitly include emittance exchange are also presented. Round-to-flat beam transform, transverse slicing, and longitudinal bunch coalescence are possible components of the alternative approach. A more explicit understanding of solenoidal cooling beam dynamics is introduced.

## INTRODUCTION

Scenarios for a high-energy high-luminosity collider require cooling the beam transversely to ~0.00003m (rms, normalized) while allowing a longitudinal emittance of ~0.1m (rms, normalized).[1] The present 6-D cooling systems cool the muons to ~0.0003m transversely and ~0.001m longitudinally.[2] Thus the collider scenarios require a "final cooling" system that reduces transverse emittances by a factor of ~10 while allowing longitudinal emittance increase. Previously, Palmer et al. have developed such a system, which includes transverse ionization cooling of low-energy muons within high field solenoids.[3, 4] At low-energies, the variation of momentum loss with energy anti-damps the beam longitudinally, increasing the longitudinal emittance, Figure 1 shows the progression of emittances throughout a collider cooling scenario, with the "final cooling" portion of that displayed as the lines with transverse emittance decrease with longitudinal emittance increase leading to final values at $\varepsilon_t$ = 25μ and $\varepsilon_L$ = ~30---60mm. More recently, Sayed et al. [5] have developed a detailed model of the final cooling system with G4Beamline tracking results that obtain performance similar to the Palmer baseline design. These systems and simulations are discussed below.

Since this "final cooling" is predominantly an emittance exchange between transverse and longitudinal dimensions, it is possible that similar results could be obtained in a final cooling system that explicitly incorporates emittance exchanges, and avoid the very large magnetic fields and very low-frequency rf with very-low-energy muons required at the end of the baseline systems. Approaches toward this are being developed by Summers et al. [6] A "round-to-flat" transform, much like that demonstrated at the Fermilab photoinjector,[7] could be used. This could be combined with a transverse beam slicer and longitudinal bunch recombiner, to obtain the small transverse emittance in both planes within a single bunch. This concept is described below, and variations which can reach the desired emittance goal with or without the round to flat transform are discussed

The "round-to-flat" transform within final cooling relies on the use of a solenoid cooling system without field flips, which then naturally develops an asymmetric emittance with one transverse circular mode damped and the other undamped. The somewhat different dynamics associated with this merits further study, and is analyzed in an intial cooling configuration.

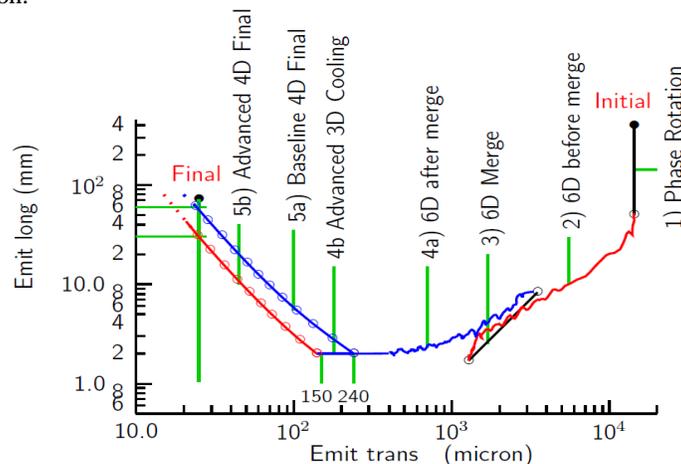

**FIGURE 1.** Overview of the evolution of emittance parameters for muon collider cooling systems. [4]

# BASELINE FINAL COOLING SYSTEM

In the baseline scenario, the μ beams are cooled as much as possible within a 6-D cooling system. The baseline 6-D cooling systems can cool the beam down to $\varepsilon_t$ ~0.0003m transversely and $\varepsilon_L$ ~0.015m. Further transverse cooling by a factor of ~2 may be possible using a non-flip ~25T solenoidal lattice, as indicated by the 150μ point in fig. 1. For final cooling, the beam momentum is reduced initially to 135 MeV/c and only transverse cooling is used.. The final cooling system consists of ~a dozen stages. Each stage consist of a high-field small bore magnet with an $H_2$ absorber within the magnet, followed by an rf and drift system within lower-field to phase-rotate and reaccelerate the muons. From stage to stage, the muon beam energy is reduced (from 66 MeV toward 5MeV) and the magnet field strength is increased to minimize $\varepsilon_t$. The relevant equations are:

$$\varepsilon_{N,eq} \cong \frac{\beta_t E_s^2}{2\beta mc^2 L_R (dE/ds)} \qquad \beta_t(m) \cong \frac{2P_\mu(GeV/c)}{0.3B(T)}$$

With B=40T and $p_\mu$=33 MeV/c ($E_\mu$ =5MeV), $\beta_t \approx$ 0.56cm and $\varepsilon_{N,eq} \approx$ 0.00001m. However, energy loss is strongly antidamping at low energies and the longitudinal emittance increases dramatically, and the final cooling lattices do not include the emitttance exchange needed to obtain longitudinal cooling. In the final stages of cooling, this antidamping is as large as the transverse damping; the 6-D emittance $\varepsilon_t^2 \varepsilon_L$ is roughly constant. In the model, the bunches are lengthened and rf rotated between absorbers to keep dp/p < ~10%. This increases the bunch length from 5cm to $\sigma_{ct}$ = 4m by end of cooling. The rf frequency must decrease correspondingly, from ~201MHz at start to ~4MHz at the end. RF frequencies < 20 MHz were considered unrealistic and the last five stages required induction linacs.

Parameters of the 14 stages of a final cooling system are displayed graphically in figure 2.

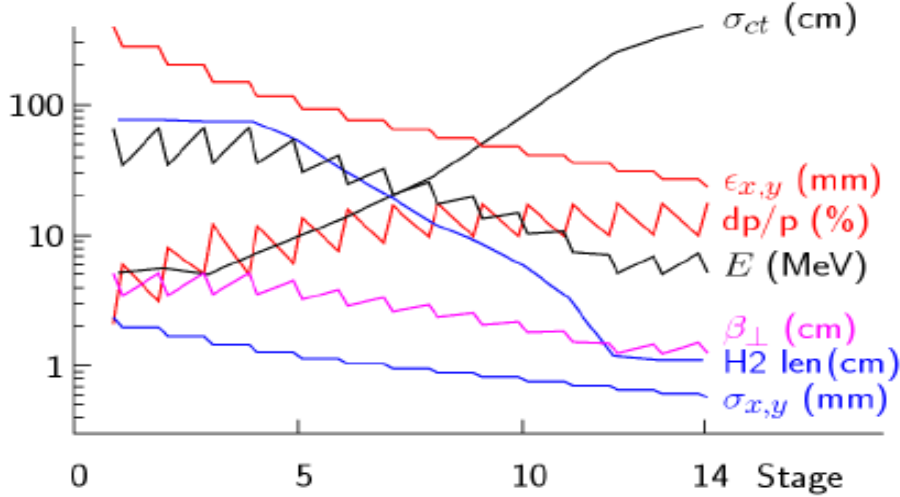

**Figure 2.** Overview of the evolution of beam and system parameters through the 14-stage final cooling system.[3]

The energy loss was simulated within ICOOL and the energy-phase motion tracked with a 1-D model, obtaining final emittances of $\varepsilon_t$ ~0.000025m transversely and $\varepsilon_L$ ~0.72m with ~33% beam loss in an~76m long system.
Major challenges in the design include the cost of high-field magnets, the low-frequency rf, and the awkward deceleration and reacceleration of low-energy μ's.

## Simulation of a Final Cooling System

H. Sayed et al. have recently simulated a final cooling system within G4Beamline, with more fully realistic models for the magnetic fields and rf systems (coils and pillbox rf cavities with Be windows) and of the matching between stages.[5] Peak magnetic fields are limited to < 30T.

In that implementation there are 16 stages with momentum decreasing from ~135 MeV/c to ~55MeV/c (13 MeV). Each stage consists of a Liquid Hydrogen absorber within a high-field solenoid followed by a drift with following rf cavities for phase-energy rotation and reacceleration. Peak magnetic field are limited to < 30T. The rf is simulated by single frequency cavities varying from 325 to 20 MHz. Some of the stages are followed by field-flips to balance the cooling between transverse degrees of freedom. While each of the stages cools transversely, the longitudinal heating/ anti-damping is as great as the transverse cooling in each stage. 6-D emittance is diluted by a factor of ~2 over the full system.

The total length of the system is 135m and the beam is cooled transversely from 300 to 55μ while the longitudinal emittance increases from ~1.5mm to 72mm. In simulation ~50% of the initial μ's are lost to apertures and decay. The performance is somewhat less than that of the baseline (transverse emittance is ~twice the desired goal), as may be expected in a first detailed simulation. Also, the simulations did not reach as extreme values in B, $f_{rf}$, and $E_\mu$ as proposed in the baseline (which went to ~40—50T, <10MHz, 5MeV). Further design optimization could improve the final parameters, particularly if more extreme fields are permitted.

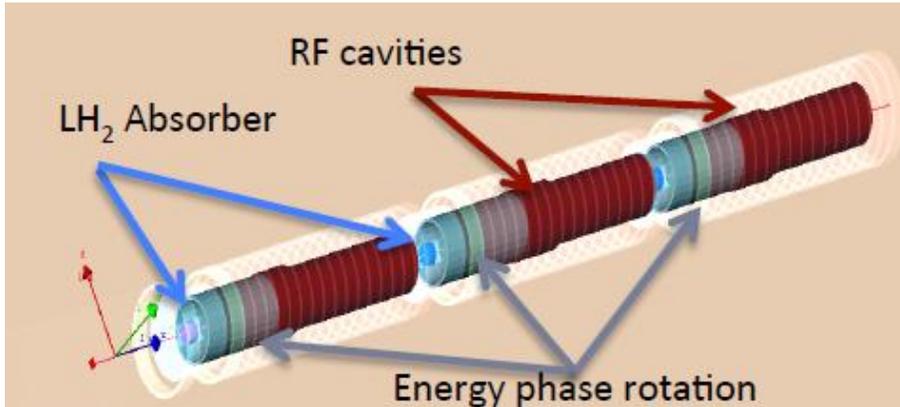

**Figure 3.** View of 3 stages in the final cooling as simulated. Each stage consists of an $H_2$ absorber within a high-field solenoid followed by a drift with bunching rf for E-φ rotation, and rf for reacceleration within ~2T fields.

## FINAL COOLING WITH BUNCH SLICING

An alternative approach to final cooling is presented by D. Summers et al.[6] Since the final cooling is dominated by emittance exchange, the approach here is to emphasize explicit emittance exchange and avoid the use of very-low frequency rf, very-low energy beams and high fields. The final cooling is envisioned as four stages:

1. Transverse Cooling. A cooling system similar to that of the baseline cooling system is used to cool the beam transversely within magnetic fields and rf systems that are relatively reasonable: $P_\mu$ = ~100MeV/c, B <25T, $f_{RF}$ > ~150 MHz. This would be much like the first 4—5 stages of the baseline system. Field-flips would not be placed between stages, enabling the development of cyclotron/drift asymmetry that can enable the round to flat transform. The length of that system should be ~40m, and it should cool $\varepsilon_t$ to ~$10^{-4}$m, while $\varepsilon_L$ → ~0.004m.
2. Round to flat beam transform. Following the technique developed for the ILC injector and other applications,[7] a solenoid → three skew- quad system transforms a "round" (large drift, small cyclotron modes)to a flat (large x, small y) emittance: $\varepsilon_t$ → $\varepsilon_x$ = 0.0004, $\varepsilon_y$ = 0.000025.
3. Transverse slicing. The beam is sliced using multiple passes through a "slow-extraction –like" septum into a string of bunches (~16). The slices are in the thicker emittance transverse plane, obtaining bunches with equal transverse emittances $\varepsilon_x$ = 0.000025, $\varepsilon_y$ = 0.000025.
4. Longitudinal recombination. The train of ~16 bunches is accelerated to a larger energy( ~10 GeV?), where a snap coalescence in a medium-energy storage ring combines these into a single bunch with enlarged longitudinal emittance ($\varepsilon_x$ = 0.000025, $\varepsilon_y$ = 0.000025, $\varepsilon_L$ =~ 0.064m).

### Discussion of method

The round-to-flat transform has been demonstrated at Fermilab, where an electron beam produced at a source within a ~0.09T magnet and accelerated to 15 MeV is transformed by skew quads to a flat beam with $\varepsilon_x/\varepsilon_y$ =~100.[7]

D. Summers et al. have simulated a round-to-flat transform at μ cooling parameters. The transform takes a thermal 115 MeV/c beam distribution at $\varepsilon_t = 10^{-4}$m within B=4T and passes it through a triplet of skew quads, obtaining a flat beam with an emittance ratio of 40. While this approximates final cooling conditions, a further simulation using beam generated from a cooling system is needed.

A longitudinal bunch combination at parameters similar than those needed here has been simulated by Johnson et al. (Their example was snap coalescence of 17 μ bunches in a 21 GeV/c storage ring.)[8]

## Final cooling without "round-to-flat"

The round-to-flat transform requires somewhat specialized matching conditions and we have not fully confirmed whether they are readily obtained in our final cooling scenario. However the present method could be readapted without that transform. The sequence would be:

1. Transverse Cooling. A cooling system similar to that of the baseline cooling system is used to cool the beam transversely within magnetic fields and rf systems that are relatively reasonable: $P_\mu = $ ~100MeV/c, B <25T, $f_{RF} > $ ~150 MHz. This would be much like the first 4—5 stages of the baseline system. Field-flips would be placed between stages to obtain symmetric x-y cooling. The length of that system should be ~40m, and it should cool $\varepsilon_x$ and $\varepsilon_y$ to ~$10^{-4}$m, while $\varepsilon_L \rightarrow$ ~0.004m.
2. Transverse slicing. The beam is sliced using multiple passes through a "slow-extraction–like" septum into a string of bunches (~10). The slices are in one plane, obtaining bunches with asymmetric transverse emittances: $\varepsilon_x = 0.00001$, $\varepsilon_y = 0.0001$.
3. Longitudinal recombination. The train of ~10 bunches is accelerated to a larger energy( ~10 GeV?), where a snap coalescence in a medium-energy storage ring combines these into a single bunch with enlarged longitudinal emittance ($\varepsilon_x = 0.00001$, $\varepsilon_y = 0.0001$, $\varepsilon_L =$~ 0.04m).
4. The beams accelerate and collide as flat beams, requiring a flat beam collision geometry, which has advantages and disadvantages with respect to round beam collisions. Luminosity with collisions of $\varepsilon_x = 0.00001$, $\varepsilon_y = 0.0001$ would be matched in luminosity to collisions of $\varepsilon_{t=}(\varepsilon_x \varepsilon_y)^{1/2}$=~0.00003m round beams.

The previous transform and slicing with round-to-flat could also be modified to end up with asymmetric flat beams, requiring flat beam collisions in a Collider. This could be advantageous.

The alternative approaches presented in this section avoid some of the more difficult components of the baseline approach (use of very high magnetic fields, very-low-frequency rf, use of very low-energy muons) and replaces them with higher-energy phase space manipulation.

## CIRCULAR MODES IN SOLENOIDAL FOCUSING

For ionization cooling at moderate to smaller energies ($P_\mu < 300$ MeV/c) the optimal focusing method is solenoidal focusing. Within solenoidal fields x and y motion is strongly coupled, and the kinetic momentum mγv is merged with the magnetic potential A to obtain the canonical momentum.

Writing the kinetic momentum as $k_i = m\gamma v_i$, the canonical momentum $p_i$ within a longitudinal solenoidal field B can be written as:

$$p_x = k_x + \frac{eB}{2c} y \quad, \quad p_y = k_y - \frac{eB}{2c} x.$$

These momenta are complementary to the horizontal and vertical positions x and y. With these coordinates, one finds that the beam angular momentum

$$\vec{L} = r \times p = (x\, p_y - y\, p_x)\, \hat{z}$$

would naturally have an expectation value of $L = -\frac{eB}{2c}\langle x^2 + y^2 \rangle$ (if <x $k_y$>=<y $k_x$>=0), and this correlation term appears in the 4-D emittance evaluation. As discussed by Kim,[9] a round beam within a solenoid would have a 4-D emittance of:

$$\varepsilon_{4D} = \varepsilon_T^2 = \varepsilon_+ \varepsilon_- = (\varepsilon_P + L)(\varepsilon_P - L)$$

Within solenoidal focusing, Burov et al.[10, 11] have noted that a more natural choice of coordinates would be the circular modes –labeled cyclotron and drift modes. The cyclotron mode coordinates are:

$$\begin{pmatrix} \kappa_1 \\ \kappa_2 \end{pmatrix} = \sqrt{\frac{c}{eB}} \begin{pmatrix} k_y \\ k_x \end{pmatrix} = \sqrt{\frac{c}{eB}} \begin{pmatrix} p_y + \frac{eB}{2c} x \\ p_x - \frac{eB}{2c} y \end{pmatrix}$$

and are simply proportional to the kinetic momentum coordinates.

The drift mode coordinates are:

$$\begin{pmatrix} \xi_1 \\ \xi_2 \end{pmatrix} = \sqrt{\frac{eB}{c}} \begin{pmatrix} d_x \\ d_y \end{pmatrix} = \sqrt{\frac{eB}{c}} \begin{pmatrix} x - \frac{c}{eB} k_y \\ y + \frac{c}{eB} k_x \end{pmatrix} = \sqrt{\frac{eB}{c}} \begin{pmatrix} \frac{x}{2} - \frac{c}{eB} p_y \\ \frac{y}{2} + \frac{c}{eB} p_x \end{pmatrix}$$

and are proportional to the centers of the Larmor motion within the solenoid field and are more closely associated with the position coordinates.

Ionization cooling damps the kinetic momentum, without affecting the position. Therefore, the cyclotron mode is damped by ionization cooling, with both coordinates damped, obtaining a partition number of 2. The drift mode is relatively undamped (partition number = 0). Over several periods of energy loss and reacceleration, one obtains a small-emittance in cyclotron mode without damping in drift mode, and a large ratio of those emittances, which are identified with $\varepsilon_-$ and $\varepsilon_+$ in ref. [11] (A factor of ~10 is readily obtained in simulations.) These modes are rotated into x and y coordinates by round-to-flat beam optics.

To obtain damping of both modes, as needed in "4-D" and "6-D" phase-space cooling, field flips (transitions from $B = B_0$ to $-B_0$) between cooling sections are required. In a single flip, the drift mode and cyclotron modes are exchanged and the previously drift mode coordinates are damped. With periodic field flips, both transverse modes are equally damped, with an average partition number of 1, obtaining 4-D cooling. (6-D cooling requires further mixing with the longitudinal motion.)

## Comparison of flip and non-flip cooling

To present a comparison between flip and non-flip cooling we consider the cooling section of the IDS neutrino factory.[12] The cooling section is an alternating solenoid lattice with a field flip every 0.75m. A non-flip lattice with similar focusing consists of a constant field B=~2T lattice. In the flip lattice, x and y emittances damp at the same rate. The initial beam angular momentum, obtained from the beam source within a solenoid, damps at a rate similar to the emittance damping, except that there is no compensating heating term.

In a non-flip cooling simulation, the initial cooling rate is the same, but all of the cooling is in the cyclotron mode ($\varepsilon_c$). The drift mode $\varepsilon_d$ does not damp and the beam size does not decrease. Canonical angular momentum increases and the emittance ratio $\varepsilon_-/\varepsilon_+$ is ~0.1 after ~75m of cooling, corresponding to the IDS front end. While the product of emittances is a bit larger, the cooled emittance $\varepsilon_-$ is ~1/2 that of $\varepsilon_x$ in the flip case. With a flat to round beam transform, one obtains a flat beam, which may be useful in some applications.

## ACKNOWLEDGMENTS

We thank M. Palmer, K. Yonehara, H. Sayed, R. Palmer, D. Summers, Y. Alexahin, and T. Hart for helpful contributions.